\documentclass{amsart}
\usepackage{amsmath}
\usepackage{amsthm,amssymb}
\usepackage{bm}
\usepackage{graphicx}
\usepackage{subfigure}
\usepackage{booktabs}
\usepackage{mathtools}
\usepackage{multirow}
\usepackage[hmargin=2cm,vmargin=2cm]{geometry}

\DeclareMathOperator*{\argmin}{arg\,min}

\DeclarePairedDelimiter\abs{\lvert}{\rvert}%
\title{Introduction to minimum message length inference}
\author{Enes Makalic and Daniel F. Schmidt}
\begin{document} 
\begin{abstract}
The aim of this manuscript is to introduce the Bayesian minimum message length principle of inductive inference to a general statistical audience that may not be familiar with information theoretic statistics. We describe two key minimum message length inference approaches and demonstrate how the principle can be used to develop a new Bayesian alternative to the frequentist $t$-test as well as new approaches to hypothesis testing for the correlation coefficient. Lastly, we compare the minimum message length approach to the closely related minimum description length principle and discuss similarities and differences between both approaches to inference.
%
\end{abstract}
\maketitle
%


%
%
\section{Minimum message length}
\label{sec:mml}
Introduced in the late 1960s by Wallace and Boulton, the minimum message length (MML) principle~\cite{WallaceBoulton68, WallaceFreeman87, WallaceDowe99a, Wallace05} is a framework for inductive inference rooted in information theory. The key idea underlying MML is that both parameter estimation and model selection problems can be interpreted as examples of data compression. It is well known that a random data string is not compressible. Conversely, if we have managed to compress a string of data we have (with high probability) learned something about the underlying structure of the data. Given data ${\bf y} \in \mathbb{R}^n$, the MML principle provides a method for computing the minimum length of a message that describes this data. In the MML approach, this message length is the single necessary inferential quantity. For the MML message to be decodable by a receiver, we require that the message encodes both a model for the data as well as the data itself. Formally, a MML message comprises two parts: 
\begin{enumerate}
\item the \emph{assertion}: describes the structure of the model, including all model parameters $\bm{\theta} \in \bm{\Theta} \in \mathbb{R}^p$. Let $I(\bm{\theta})$ denote the codelength of the assertion.
\item the \emph{detail}: describes the data ${\bf y}$ using the model $p({\bf y} | \bm{\theta})$ nominated in the assertion. Let $I({\bf y} | \bm{\theta})$ denote the codelength of the detail.
\end{enumerate}
The total length of the MML message, $I({\bf y}, \bm{\theta})$, measured in units of information (for example, bits) is the sum of the lengths of the assertion and the detail: 
\begin{equation}
\label{eqn:mml:codelength}
I(D, \bm{\theta}) = \underbrace{ I(\bm{\theta}) }_{\rm assertion} + \underbrace{I({\bf y} | \bm{\theta})}_{\rm detail} .
\end{equation}
The length of the assertion measures the complexity of the model, with longer assertions able to state more parameters with high accuracy or describe more complicated model structures. In contrast, a short assertion may encode the model parameters imprecisely and describe only simple models. The length of the detail tells us how well the model stated in the assertion is able to fit or describe the data. A complex model, that is one with a long assertion, will have lots of explanatory power and be able to encode more data strings using fewer bits compared to a simple model. MML seeks the model
\begin{equation}
\hat{\bm{\theta}}({\bf y}) = \argmin_{\bm{\theta} \in \bm{\Theta}} \left\{ I({\bf y}, \bm{\theta}) \right\}
\label{eqn:mml}
\end{equation}
that minimises the length of the two-part message message. The key point is that minimising the two part message requires balancing the complexity of the model (assertion) with how well the model describes the data (detail). Ideally, we wish to find the simplest model that fits the observed data well enough; essentially, a formalisation of the famous razor of Occam. An advantage of MML is that the message length, measured in (say) bits, is a universal gauge that allows comparison across models with different model structures and numbers of parameters. As long as we can compute the MML codelengths of models, we can compare them. In this fashion, an MML practitioner is able to compare, for example, a linear regression model~\cite{SchmidtMakalic09c}, to a finite mixture model~\cite{WallaceDowe00} to a decision tree~\cite{WallacePatrick93} via their codelengths for some observed data set.

The aim of this article is to introduce the MML principle of inductive inference to readers that may not be overly familiar with information theoretic statistics. We will examine two key MML codelength solutions to (\ref{eqn:mml}): the exact, or the so called Strict MML codelength (see Section~\ref{sec:smml}), and a practical approximation to the exact codelength (see Section~\ref{sec:mml87}). We will demonstrate how to perform MML analysis of two common statistical problems: the two-sample $t$-test problem (see Section~\ref{sec:mml:ttest}) and testing of the correlation coefficient (see Section~\ref{sec:mml:corr}). Lastly, we  compare MML to minimum description length (MDL), an alternative approach to inductive inference developed independently by Rissanen and colleagues~\cite{Rissanen78,Rissanen96,RissanenRoos07a,Rissanen09b}.


%
%
\subsection{Strict Minimum Message Length}
\label{sec:smml}
%
The exact solution to (\ref{eqn:mml}) is usually called Strict MML~\cite{WallaceBoulton75, Wallace05}, and may be viewed as a ``gold standard' which more tractable formulae seek to approximate. To describe Strict MML we shall closely follow the notation in~\cite{Wallace05} (pp. 154--155). Let
\begin{equation}
	r({\bf y}) = \int \pi(\bm{\theta}) p({\bf y} | \bm{\theta}) \, d\bm{\theta} ,
\end{equation}
denote the marginal distribution of the $n$ data points ${\bf y}$ given a prior distribution $\pi(\bm{\theta})$ over the model parameters $\bm{\theta} \in \mathbb{R}^p$. We note that MML is strictly Bayesian and requires a prior distribution for the model parameters in order to perform inference. Next we define the subset $\Theta^* = \{\bm{\theta}^*_1, \bm{\theta}^*_2, \ldots\} \subset \Theta$ as the set of parameters that can be used to encode the data ${\bf y}$, where each $\bm{\theta}_j^*$ may be encoded with different codelength. Recall that the assertion describes the model structure and all model parameters $\bm{\theta} \in \Theta$. To encode the assertion we require that $\Theta^*$ be a countable set as the assertion is finite and can encode only a countable subset of $\Theta$. Clearly, in many inference problems the parameter set $\Theta$ is continuous and not countable and the question of how to quantize the parameter space $\Theta$ into a countable set $\Theta^*$ is a key issue in MML inference. Assuming we have somehow obtained the quantized parameter set $\Theta^*$, the code for $\Theta^*$ implies a probability distribution over the set $\Theta^*$ since
\begin{equation*}
    p(\bm{\theta}^*_j) = q_j > 0, \quad \sum_{j: \bm{\theta}_j^* \in \Theta^*} q_j = 1, \quad (j = 1,2,\ldots).
\end{equation*}
The assertion is completely determine by the set $\Theta^*$ and a distribution (code) over the set implying that a $\bm{\theta}_j^*$ should be optimally encoded with length $-\log q_j$.

The second part of the message, the detail, is the encoding of the data given the model $\hat{\bm{\theta}} (\equiv \bm{\theta}_j^*)$ specified in the assertion. Once we have chosen the model $\hat{\bm{\theta}}$, the data ${\bf y}$ can be encoded with length $-\log p({\bf y} | \hat{\bm{\theta}})$. Let $t_j = \{{\bf y} \in \mathbb{R}^n: m({\bf y}) = \bm{\theta}^*_j\}$ denote the set of data values ${\bf y}$ that will be encoded using some parameter $\bm{\theta}^*_j \in \Theta^*$ in the detail component of the message. Similarly, denote by $q_j = q(\bm{\theta}^*_j)$ the probability mass associated with parameter $\bm{\theta}^*_j \in \Theta^*$ used to construct the assertion codeword for $\bm{\theta}^*_j$. Strict MML seeks the mapping $m(\cdot)$ which minimizes the expected codelength of the message~\cite{Wallace05} (pp. 155)
\begin{align}
	I_S &= \sum_{{\bf y} \in \mathbb{R}^n} r({\bf y}) \left[ - \log q(m({\bf y})) - \log p({\bf y} | m({\bf y}) \right], 
	\label{eqn:smml:codelength1}
\end{align}
where the expectation is taken with respect to the marginal distribution of the data. The first term in the expectation, $-\log q(m({\bf y}))$, is the length of the codeword stating parameter $\bm{\theta}^*_j \in \Theta^*$ in the assertion, while the second term, $-\log p({\bf y} | m({\bf y}))$, denotes the codelength of a data set ${\bf y}$ encoded using  parameter $\bm{\theta}^*_j$.

\begin{table*}[tb]
\footnotesize
\begin{center}
\begin{tabular}{cllc} 
\toprule[1pt]
$n$ & SMML PARTITION & $\Theta^*$ & $I_S$ (bits) \\
\cmidrule{1-4}
  1    &\{0..1\}                                          &\{0.500\}                                         & 1.000\\
  2    &\{0..0, 1..2\}                                    &\{0.000, 0.750\}                                  & 1.667\\
  3    &\{0..0, 1..3\}                                    &\{0.000, 0.667\}                                  & 2.085\\
  4    &\{0..0, 1..3, 4..4\}                              &\{0.000, 0.500, 1.000\}                           & 2.454\\
  5    &\{0..0, 1..4, 5..5\}                              &\{0.000, 0.500, 1.000\}                           & 2.704\\
  6    &\{0..0, 1..5, 6..6\}                              &\{0.000, 0.500, 1.000\}                           & 2.962\\
  7    &\{0..3, 4..7\}                                    &\{0.214, 0.786\}                                  & 3.165\\
  8    &\{0..2, 3..7, 8..8\}                              &\{0.125, 0.625, 1.000\}                           & 3.337\\
  9    &\{0..0, 1..5, 6..9\}                              &\{0.000, 0.333, 0.833\}                           & 3.491\\
 10    &\{0..0, 1..4, 5..9, 10..10\}                      &\{0.000, 0.250, 0.700, 1.000\}                    & 3.647\\
 11    &\{0..0, 1..5, 6..10, 11..11\}                     &\{0.000, 0.273, 0.727, 1.000\}                    & 3.762\\
 12    &\{0..0, 1..6, 7..11, 12..12\}                     &\{0.000, 0.292, 0.750, 1.000\}                    & 3.887\\
 13    &\{0..0, 1..6, 7..12, 13..13\}                     &\{0.000, 0.269, 0.731, 1.000\}                    & 3.998\\
 14    &\{0..3, 4..10, 11..14\}                           &\{0.107, 0.500, 0.893\}                           & 4.107\\
 15    &\{0..0, 1..5, 6..12, 13..15\}                     &\{0.000, 0.200, 0.600, 0.933\}                    & 4.204\\
 16    &\{0..0, 1..5, 6..12, 13..16\}                     &\{0.000, 0.188, 0.563, 0.906\}                    & 4.289\\
 17    &\{0..0, 1..6, 7..13, 14..17\}                     &\{0.000, 0.206, 0.588, 0.912\}                    & 4.372\\
 18    &\{0..0, 1..5, 6..12, 13..17, 18..18\}             &\{0.000, 0.167, 0.500, 0.833, 1.000\}             & 4.457\\
 19    &\{0..0, 1..5, 6..12, 13..18, 19..19\}             &\{0.000, 0.158, 0.474, 0.816, 1.000\}             & 4.531\\
 20    &\{0..0, 1..6, 7..14, 15..19, 20..20\}             &\{0.000, 0.175, 0.525, 0.850, 1.000\}             & 4.601\\
 21    &\{0..0, 1..6, 7..14, 15..20, 21..21\}             &\{0.000, 0.167, 0.500, 0.833, 1.000\}             & 4.665\\
 22    &\{0..0, 1..6, 7..15, 16..21, 22..22\}             &\{0.000, 0.159, 0.500, 0.841, 1.000\}             & 4.737\\
 23    &\{0..3, 4..11, 12..19, 20..23\}                   &\{0.065, 0.326, 0.674, 0.935\}                    & 4.801\\
 24    &\{0..2, 3..9, 10..17, 18..23, 24..24\}            &\{0.042, 0.250, 0.563, 0.854, 1.000\}             & 4.863\\
 25    &\{0..0, 1..6, 7..15, 16..22, 23..25\}             &\{0.000, 0.140, 0.440, 0.760, 0.960\}             & 4.920\\
 26    &\{0..0, 1..6, 7..14, 15..22, 23..26\}             &\{0.000, 0.135, 0.404, 0.712, 0.942\}             & 4.974\\
 27    &\{0..0, 1..6, 7..15, 16..23, 24..27\}             &\{0.000, 0.130, 0.407, 0.722, 0.944\}             & 5.025\\
 28    &\{0..3, 4..12, 13..21, 22..27, 28..28\}           &\{0.054, 0.286, 0.607, 0.875, 1.000\}             & 5.080\\
 29    &\{0..4, 5..13, 14..22, 23..28, 29..29\}           &\{0.069, 0.310, 0.621, 0.879, 1.000\}             & 5.129\\
 30    &\{0..0, 1..5, 6..14, 15..23, 24..29, 30..30\}     &\{0.000, 0.100, 0.333, 0.633, 0.883, 1.000\}      & 5.176\\
\vspace{-3mm} \\ 
\bottomrule[1pt]
\vspace{+1mm}
\end{tabular}
\caption{SMML partitions for a binomial distribution under a uniform prior on the probability of success. The countable set $\Theta^*$ denotes the parameters that would be used to code the data.\label{tab:smml:binomial}}
\end{center}
\end{table*}

First, note that (\ref{eqn:smml:codelength1}) minimizes the expected codelength over all possible data values because the map $m(\cdot)$ depends on unobserved data values that could be generated by $r({\bf y})$, and not just on the particular data set that has been observed. Second, the expression for the expected codelength includes a sum over all possible data sets of size $n$ as SMML considers the space of data to be countable. This is reasonable since we are interested in computing the length of a message that  transmits the observed data and this message must be finite. The expected codelength can be written as
\begin{align}
\nonumber
I_S &= -\sum_{\bm{\theta}^*_j \in \Theta^*} \sum_{{\bf y}_i \in t_j} r({\bf y}_i)  \left( \log q(\bm{\theta}^*_j) + \log p({\bf y}_i | \bm{\theta}^*_j) \right),\\
\label{eq:smml:optim}
&= \underbrace{-\sum_{\bm{\theta}^*_j \in \Theta^*} q(\bm{\theta}_j) \log q(\bm{\theta}^*_j) }_{\text{assertion}}  \underbrace{-\sum_{\bm{\theta}^*_j \in \Theta^*} \sum_{y_i \in t_j} r({\bf y}_i) \log p({\bf y}_i | \bm{\theta}^*_j) }_{\text{detail}},
\end{align}
where the coding probability of $\bm{\theta}^*_j \in \Theta^*$ is 
\begin{equation}
	q(\bm{\theta}^*_j) = \sum_{{\bf y}_i \in t_j} r({\bf y}_i) ,
\end{equation}
which is the sum of the marginal probabilities of the data values that result in the estimate $\bm{\theta}^*_j\in \Theta^*$ being used in the detail. The first term in the expected Strict MML codelength is the \emph{expected} length of the assertion, while the second term gives the expected length of the detail. Strict MML seeks to minimize the expected codelength since the map $m(\cdot)$ depends on all data values ${\bf y} \in \mathbb{R}^n$ and not just the observed data.

By seeking the optimal map $m(\cdot)$, Strict MML partitions the data space into disjoint regions $t_j$, such that all data sets that fall within a particular region $t_j$ are encoded with the same parameter value $\bm{\theta}^*_j \in \Theta^*$. This partitioning of the data space yields the quantized parameter set $\Theta^*$ where members of $\Theta^*$ are the feasible parameter estimates, with $m({\bf y})$ being the point estimate associated with data string ${\bf y}$.  

The countable set $\Theta^*$ that minimises (\ref{eq:smml:optim}) is not necessarily unique; however this is of no practical consequence as all optimal $\Theta^*$ result in an efficient encoding, and hence feasible parameter values~\cite{Wallace05} (pp. 161). The Strict MML codelength is invariant to data and model representation and represents the shortest two-part codelength for a message that conveys the data and a model for the data. Strict MML has very close links to fundamental concepts such as Turing machines and Kolmolgorov complexity~\cite{WallaceDowe99c}. We note that, while as have assumed the data space to be discrete in our exposition of Strict MML, this is not necessary and the Strict MML code for continuous data is a straightforward extension of the codelength (\ref{eq:smml:optim}); see~\cite{Wallace05}, pp. 166--169. 

Unfortunately, the computational complexity of partitioning the data space and obtaining $\Theta^*$ is NP complete in general and only practically computable for simple models~\cite{FarrWallace02}. To see this, we note that the Strict MML data space partitioning problem is an example of the well-known set partitioning problem (also known as the coalition structure generation problem in the artificial intelligence research community) provided the data space is countably finite~\cite{LamarchePerrin14,RahwanEtAl15}. The set partitioning problem is known to be NP complete in general and cannot be solved by brute-force search as the total number of possible partitions, assuming a finite data space of size $n$, is the $n$-th Bell number $B_n$~\cite{Bell34} and satisfies $\alpha n^n \leq B_n \leq n^n$ for some $\alpha > 0$~\cite{SandholmEtAl99}. In some cases, constraints that impose structure and reduce the search space can be assumed. For example, if the data space is one-dimensional and ordered like in the example below, this is know as the ordered set partitioning problem and is identical to the shortest path problem that can be solved in $O(n^2)$ polynomial time. \\

{\noindent \bf Example: } Consider data $y$ representing the number of successes in $n$ trials which follows the binomial distribution with unknown success probability $\theta \in \Theta = [0,1]$; that is,
\begin{equation}
\label{eqn:pdf:binomial}
	p(y | \theta) = \binom{n}{y} \theta^y (1 - \theta)^{n - y} .
\end{equation}
Suppose that the success probability $\theta$ is assumed  equally likely to take on any value in the parameter space $\Theta$ a priori; in other words, we assume the uninformative prior distribution $\pi(\theta) = 1$. The marginal distribution of the data is
\begin{equation*}
    r(y) = \int_0^1 p(y | \theta) \pi(\theta) d\theta = \int_0^1 \binom{n}{y} \theta^y (1 - \theta)^{n - y} d\theta = \frac{1}{n+1} .
\end{equation*}
To construct a Strict MML code for the binomial distribution we require the map $m(\cdot)$ that minimizes the expected codelength (\ref{eq:smml:optim}) over all possible data sets $y \in \mathbb{N} = \{0,1,\ldots,n\}$, where the expectation is taken with respect to $r(y)$. This optimal codelength will be based on an optimal partitioning of the data space which implicitly yields the quantized parameter set $\Theta^*$. The expected Strict MML codelength for the binomial distribution is
\begin{equation*}
    I_S = -\sum_{\theta^*_j \in \Theta^*} q(\theta_j^*) \log q(\theta^*_j)  -\sum_{\theta^*_j \in \Theta^*} \sum_{y_i \in t_j} \left(\frac{1}{n+1} \right) \log \binom{n}{y_i} (\theta_j^*)^{y_i} (1 - \theta_j^*)^{n - y_i} , \quad 
	q(\theta^*_j) = \sum_{y_i \in t_j} \frac{1}{n+1},
\end{equation*}
and $t_j$ is the set of data values $y_i$ that will be encoded using $\theta_j^* \in \Theta^*$. For a given partition $\Theta^*$, the contribution each estimate $\theta_j^*$ makes to the expected codelength is part of the detail only and does not influence the codelength of the assertion. To minimize the expected codelength for a particular $\theta_j^*$, we therefore only need to minimize the second part of the message
\begin{equation*}
    - \sum_{y_i \in t_j} r(y_i) \log \binom{n}{y_i} (\theta_j^*)^{y_i} (1 - \theta_j^*)^{n - y_i} .
\end{equation*}
Clearly, the $\hat{\theta}^*_j$ that minimizes the above is
\begin{equation*}
    \hat{\theta}_j^* = \frac{\sum_{y_i \in t_j} r(y_i) y_i}{n \sum_{y_i \in t_j} r(y_i)} = \frac{1}{n |t_j|} \sum_{y_i \in t_j} y_i, \quad j = 1,2,\ldots ,
\end{equation*}
where $|t_j|$ denotes the size of the set $t_j$; i.e., the number of data values that are encoded with parameter $\theta_j^*$. 

It remains to determine the optimal data space partitions $t_j$ and hence the quantized set $\Theta^*$. Before we look at the optimal Strict MML partitioning, we shall look at two  examples with suboptimal codelengths. The simplest possible partitioning of the data space $\{0,1,\ldots, n\}$ assumes that all data values are in the same partition, say, $t_1 = \{0, 1, \ldots, n\}$, where the size of the partition is $|t_1| = (n+1)$. This implies that the quantized parameter set $\Theta^* = \{ \theta^*\}$ consists of a single member
\begin{equation*}
    \hat{\theta}^* = \frac{1}{n |t_1|} \sum_{y_i \in t_1} y_i = \frac{1}{2} .
\end{equation*}
which has probability $q(\theta^*) = 1$. The expected codelength of the assertion is then 0 bits and the expected codelength of the detail, and hence the total codelength, is
\begin{equation*}
        I_S = n \log 2 - \left(\frac{1}{n+1}\right)\sum_{i=0}^n  \log \binom{n}{i}  .
\end{equation*}
For example, for $n = 10$ and assuming a single partition that contains all possible data of size $n$, the Strict MML codelength for some observed data $y$ is approximately 5.01 bits; here, the estimate of the success probability is $\hat{\theta} = 0.5$. 

Alternatively, consider the more complex partitioning where the data space is divided into $(n+1)$ partitions with each  datum $y_i \in \mathbb{N}$ belonging to its own partition. In this case, there are $(n+1)$ possible estimates and $\Theta^* = \{ \theta^*_j: \theta^*_j = j / n, j=0,1,\ldots,n\}$. The probability of each estimate is 
\begin{equation*}
    q_j = q(\theta^*_j) = \sum_{y_i \in t_j} r(y_i) = \frac{1}{n+1}, \quad j = 0, \ldots, n.
\end{equation*}
The expected Strict MML codelength is now
\begin{equation*}
    I_S = \log(n+1) -\frac{1}{n+1}\sum _{i=0}^n \log \binom{n}{i} \left(\frac{i}{n}\right)^i \left(1-\frac{i}{n}\right)^{n-i} .
\end{equation*}
For the $n=10$ example, the total codelength of this partitioning is approximately 9.84 bits. This codelength is significantly worse than the single partition Strict MML code.

It is possible to compute the optimal Strict MML solution for the binomial distribution using the dynamic programming algorithm of Farr and Wallace~\cite{FarrWallace02}. The results of the algorithm are shown in Table~\ref{tab:smml:binomial} for all $1 \leq n \leq 30$; a MATLAB implementation of this algorithm can be downloaded from the Mathworks File Exchange website (ID: 80167). When $n=10$, for example, Strict MML partitions the data space into four partitions $\{0, 1..4, 5..9, 10\}$ yielding the parameter set $\Theta^* = \{ 0.00, 0.25, 0.70, 1.00 \}$. This implies that if we observe between 5 and 9 success counts, the estimated probability of success is 0.7. In contrast, if $0$ or $10$ success counts were observed, we would estimate the probability of success as 0 and 1, respectively. The Farr and Wallace algorithm is applicable to any one dimensional Strict MML problem with discrete data and is of polynomial complexity in terms of the sample size $n$.  \\

An algorithm to obtain optimal Strict MML codelengths for one dimensional continuous data from exponential families with continuous sufficient statistics was developed by Dowty~\cite{Dowty15a}. Unfortunately, there exists no general algorithm for computing optimal Stict MML partitions outside of the one dimensional (continuous or discrete) data setting. In the case of the $k$-nomial distribution with $k>2$, the Strict MML data partitioning problem is an instance of the set partitioning (coalition structure generation) over graphs. While a dynamic programming algorithm exists to solve this problem, it runs in exponential time and is only practical for small sample sizes~\cite{MichalakEtAl16}. We note that many  anytime exact algorithms and heuristic algorithms have also been proposed for this setting; see Rahwan et. al for a recent survey~\cite{RahwanEtAl15}.
\subsection{A practical approximation to the Strict MML codelength}
\label{sec:mml87}
Due to the high computational complexity of deriving the exact Strict MML codelength, Strict MML is mostly of interest from a theoretical standpoint. There exist several approximations to the Strict MML codelength with the MML87 approximation~\cite{WallaceFreeman87,Wallace05} being the most popular. Under some regularity conditions~\cite{Wallace05}) (pp. 226), the MML87 codelength for data $D$ is
\begin{equation}
\label{eqn:mml87:codelength}
	\mathcal{I}_{87}({\bf y}, \bm{\theta}) = \underbrace{-\log \pi(\bm{\theta}) + \frac{1}{2} \log \abs{J_{\bm{\theta}}(\bm{\theta})} + \frac{p}{2} \log \kappa_p}_{\rm assertion} + \underbrace{\frac{p}{2} - \log p({\bf y}|\bm{\theta})}_{\rm detail}
\end{equation}
where $\pi_{\bm{\theta}}(\bm{\theta})$ is the prior distribution for the parameters $\bm{\theta}$, $\abs{J_{\bm{\theta}}(\bm{\theta})}$ is the determinant of the expected Fisher information matrix, $p({\bf y}|\bm{\theta})$ is the likelihood function of the model and $\kappa_p$ is a quantization constant~\cite{ConwaySloane98,AgrellEriksson98}; for small $p$ we have 
\begin{equation}
\kappa_1 = \frac{1}{12}, \quad \kappa_2 = \frac{5}{36 \sqrt{3}}, \quad \kappa_3 = \frac{19}{192 \times  2^{1/3}},
\end{equation}
while, for large $p$, $\kappa_p$ is well-approximated by~\cite{Wallace05}:
\begin{equation}
\frac{p}{2} (\log \kappa_p + 1) \approx -\frac{p}{2} \log 2\pi + \frac{1}{2} \log p \pi - \gamma,
\end{equation}
where $\gamma \approx 0.5772$ is the Euler--Mascheroni constant. 

In contrast to SMML, which quantizes the parameter space by associating a single parameter estimate with a number of different data strings, the MML87 approximation achieves computational tractability by instead associating a set of models, deemed ``indistinguishable'' in an information sense, with the observed data string ${\bf y}$. This set of parameter estimates can be thought of as the range of possible parameter estimates that a plausible choice of $m(\cdot)$ would associate with the data string ${\bf y}$. The volume of this set of models, which is called the \emph{uncertainty region} in MML parlance, explicitly determines the accuracy to which the continuous model parameters should be stated (i.e., the degree of quantization). We can re-write the codelength of the assertion in terms of the volume of the uncertainty region $w(\bm{\theta})$ as follows
\begin{equation}
    \underbrace{-\log \pi(\bm{\theta}) w(\bm{\theta})}_{\rm assertion}, \quad w(\bm{\theta}) = \left( \abs{J_{\bm{\theta}}(\bm{\theta})} \kappa_p^p \right)^{-1/2} .
\end{equation}
It is clear  that the volume of the uncertainty region depends on the variation of the negative log-likelihood function. If a small change in $\bm{\theta}$ results in a large change in the log-likelihood, the volume of the uncertainty region will be relatively small, implying that the parameters must be stated with higher precision. In contrast, if the negative log-likelihood is not particularly sensitive to  small changes in $\bm{\theta}$, the volume of the uncertainty region will be fairly large. Under suitable regularity conditions, the volume of the uncertainty region is $O(n^{-p/2})$. 

For many sufficiently well-behaved models, the MML87 codelength is virtually identical to the Strict MML codelength while being simpler to compute, requiring only the prior distribution for the model parameters and the determinant of the expected Fisher information matrix. Unlike the Strict MML codelength $I_S$ (\ref{eqn:smml:codelength1}) which must be found via minimization of an expectation, taken over all possible data with respect to the marginal distribution $r(\cdot)$, the MML87 codelength is computed using the observed data only. For large sample sizes $n \to \infty$, it is easy to show that the MML87 codelength is asymptotically equivalent to the well-known Bayesian information criterion (BIC)~\cite{Schwarz78}  
\begin{equation}
\label{eqn:mml87:bic}
	\mathcal{I}_{87}({\bf y}, \bm{\theta}) = - \log p({\bf y}|\bm{\theta}) + \frac{p}{2} + O(1) ,
\end{equation}
where the $O(1)$ term depends on the prior distribution, the Fisher information and the number of parameters $p$. In fact, as the MML87 codelength $\mathcal{I}_{87}({\bf y}, \bm{\theta})$ can be interpreted as the negative logarithm of the posterior probability mass attached to a dataset ${\bf y}$ and model $\bm{\theta}$, the difference in message lengths admits the interpretation as the logarithm of the posterior-odds of two models, e.g.,
\[
    \mathcal{I}_{87}({\bf y}, \bm{\theta}_0) - \mathcal{I}_{87}({\bf y}, \bm{\theta}_1)
\]
can be interpretated as the posterior log-odds in favour of model $\bm{\theta}_1$ against $\bm{\theta}_0$. The MML87 codelength results in estimates that are invariant under (smooth) one-to-one reparametarization, just like the maximum likelihood estimate. MML87 has been applied to a wide range of statistical models including decision tress~\cite{WallacePatrick93}, factor analysis~\cite{WallaceFreeman92} and mixture models~\cite{WallaceDowe00}. \\

\noindent {\bf Example:} We will compute the MML87 codelength for the binomial distribution and compare it to the Strict MML code~\cite{Wallace05} (pp. 246). Let $y \in [0, n]$ be the count of successes which follows a binomial distribution with probability density function (\ref{eqn:pdf:binomial}) where $0 < \theta < 1$ is the probability of observing a success. Suppose we assume a uniform prior distribution for the success probability $\pi(\theta) = 1$. The determinant of the expected Fisher information matrix for a binomial distribution is $J(\theta) = n/(\theta(1-\theta))$. The volume of the uncertainty region is 
\begin{equation*}
    w(\theta) =  \left(\frac{12 (1-\theta ) \theta }{n}\right)^{1/2}
\end{equation*}
which gets smaller as $n \to \infty$, or as the success probability $\theta$ approaches $\theta \to 0$ or $\theta \to 1$.
Substituting into the MML87 codelength (\ref{eqn:mml87:codelength}) we obtain
\begin{align}
	\mathcal{I}_{87}(y, \theta) &= \frac{1}{2} \log \left( \frac{n}{\theta(1-\theta)} \right) + \frac{1}{2} \log \frac{1}{12} - \log \binom{n}{y} \theta^y (1 - \theta)^{n - y} + \frac{1}{2} \nonumber  \\
&= -\left(y + \frac{1}{2}\right) \log \theta - \left(n - y + \frac{1}{2}\right) \log (1 - \theta) + \frac{1}{2} \left(1 + \log \frac{n}{12 \binom{n}{y}^2}\right)
\end{align} 
This codelength is minimized by choosing $\theta$ to be the MML87 estimate
\begin{equation}
	\hat{\theta}(y) = \frac{y + 1/2}{n + 1} ,
\end{equation}
which results in the optimal MML87 codelength for data $y$ being
\begin{align*}
	\mathcal{I}_{87}(y)= -\left(y + \frac{1}{2}\right) \log \hat{\theta}(y) - \left(n - y + \frac{1}{2}\right) \log (1 - \hat{\theta}(y)) + \frac{1}{2} \left(1 + \log \frac{n}{12 \binom{n}{y}^2}\right) .
\end{align*} 
As an example, if $n = 10$ and $y = 3$, the MML87 estimate of the success probability is $\hat{\theta}(y) = 7/22 \approx 0.32$, the volume of the uncertainty region is $w(\theta) \approx 0.51$ and the MML87 codelength is approximately 3.61 bits. In contrast, the SMML estimate of the success probability is $\hat{\theta}(y) = 0.25$ resulting in a codelength of 3.647 bits. We can empirically show that the expected MML87 codelength with respect to the marginal distribution $r(y) = 1/(n+1)$ falls within 0.1 bits of the expected Strict MML codelength for all $n \geq 5$, while being significantly simpler to compute. 


%

%
%
\section{Minimum message length $t$-test}
\label{sec:mml:ttest}
Perhaps the most popular hypothesis test in practice is the frequentist two-sample $t$-test for comparison of two means. As an example of the $t$-test popularity, a search for the term `paired $t$-tests' returns approximately 12,000 articles from 1990 onwards in PubMed. Similarly, articles published in peer reviewed psychology journals feature, on average, more than three $t$-tests per article~\cite{WetzelsEtAl11}. Formally, the $t$-test assumes that we observe normally distributed data from two, possibly different, populations:
\begin{equation}
	Y_{1i} \sim \mathcal{N}\left(\mu + \frac{\sigma \delta}{2}, \sigma^2\right), \quad Y_{2j} \sim \mathcal{N}\left(\mu - \frac{\sigma \delta}{2}, \sigma^2\right) ,
\end{equation}
for $i=1,\ldots,n_1$ and $j = 1, \ldots, n_2$. Here, the unknown parameter $\mu \in \mathbb{R}$ is the overall (grand) mean, $\sigma > 0$ is a standard deviation that is common to both groups, $\delta>0$ is the standardised effect size and $(n_1, n_2)$ are the sample sizes of the two groups.  This formulation of the $t$-test in terms of the overall mean and standardised effect size is due to \cite{GonenEtAl05}. The usual null hypothesis says that the mean is the same in both groups while the alternative hypothesis specifies a different mean for each group; formally, we have
\begin{equation}
	\mathcal{H}_0: \delta = 0, \quad \mathcal{H}_1: \delta \neq 0 .
\end{equation}
Let $\bm{\theta}_0 = \{\mu, \sigma\} \in (\mathbb{R} \times \mathbb{R}_+) \equiv \Theta_0$ and $\bm{\theta}_1 = \{\mu, \sigma, \delta\} \in (\mathbb{R}^2 \times \mathbb{R}_+) \equiv \Theta_1$ denote the parameter set under the two models. Given data ${\bf y} = \{{\bf y}_1, {\bf y}_2\}$ where ${\bf y}_1 = (y_{1,1}, \ldots, y_{n_1,1})^\prime$ and ${\bf y}_2 = (y_{2,1}, \ldots, y_{n_2,1})^\prime$, the task is estimate the effect size $\delta > 0$ and thus determine whether the population means of the two groups are equal ($\delta = 0$).

Recently, there has been a great deal of interest in Bayesian alternatives to the frequentist $t$-test~\cite{GonenEtAl05,RouderEtAl09,WangLiu16,GronauEtAl19,Kelter21}. A key quantity in Bayesian analysis is the Bayes factor
\begin{equation}
\underbrace{ \frac{P(\mathcal{H}_1 | {\bf y})}{P(\mathcal{H}_0 | {\bf y})} }_{\text{Posterior odds}} = \underbrace{ \frac{p({\bf y} | \mathcal{H}_1)}{p({\bf y} | \mathcal{H}_0)} }_{\text{Bayes factor}} \underbrace{ \frac{P(\mathcal{H}_1)}{P(\mathcal{H}_0} }_{\text{Prior odds}} ,
\end{equation}
which is the ratio of the marginal likelihoods of the data obtained by integrating the parameters out with respect to their prior distribution; that is, 
\begin{equation}
p({\bf y} | \mathcal{H}_i) = \int p({\bf y} | \bm{\theta}_i) \pi_i(\bm{\theta}_i) \, d\bm{\theta}, \quad i = 0, 1 ,
\end{equation}
where $\pi_i(\bm{\theta}_i)$ is the prior distribution of the parameters under model $i$. The Bayes factor, henceforth denoted by BF$_{10}$, measures the amount of evidence in support of hypothesis $\mathcal{H}_1$ over hypothesis $\mathcal{H}_0$, where BF$_{10} > 3$ indicates substantial evidence in favour of $\mathcal{H}_1$~\cite{KassRaftery95}. Given a proper prior $\pi(\delta)$ on the effect size $\delta$, Gronau et al.~\cite{GronauEtAl19}, generalizing the work of \cite{GonenEtAl05,RouderEtAl09}, showed that the Bayes factor for the $t$-test problem can be expressed as 
\begin{equation}
	\text{BF}_{10} = \frac{\int T_\nu(t | \sqrt{n_\delta} \delta ) \pi(\delta)  \, d\delta }{T_{\nu}(t)}
\end{equation}
where $t$ is the observed $t$-statistic
\begin{equation}
	t = \sqrt{n_\delta} (\bar{\bf y}_1 - \bar{\bf y}_2) / s_p, \quad \nu s_p = (n_1 - 1) s_1^2 + (n_2 - 1) s_2^2, 
\end{equation}
and $s_j$ ($j=1,2)$ are the usual unbiased sample variance estimates. Here, $\nu = n_1 + n_2 - 2$ denotes the degrees of freedom, $n_\delta = (1/n_1+1/n_2)^{-1}$ is the effective sample size, and $T_\nu(\cdot|a)$ is the non-central Student $t$ distribution with degrees of freedom $\nu$ and non-centrality parameter $a$. For most prior distributions $\pi(\delta)$, the integral in the Bayes factor must be computed by numerical integration.

In order to perform the two-sample $t$ test within the MML framework, we require the codelengths of the data under the null distribution $(\delta = 0)$ and under the alternative hypothesis $(\delta \neq 0)$. Under the null model, the negative log-likelihood of the data ${\bf y}$ is
\begin{eqnarray}
	-\log \ell_0(\bm{\theta}_0) &=& \frac{n_1}{2} \log(2 \pi \sigma^2) + \frac{1}{2 \sigma^2} \sum_{i=1}^{n_1} \left(y_{1i} - \mu\right)^2 + \frac{n_2}{2} \log(2 \pi \sigma^2) + \frac{1}{2 \sigma^2} \sum_{j=1}^{n_2} \left(y_{2j} - \mu\right)^2,	\nonumber \\
	&=& \frac{n}{2} \log(2 \pi \sigma^2) + \frac{1}{2 \sigma^2} \sum_{i=1}^{n} \left(y_{i} - \mu\right)^2, \nonumber
\end{eqnarray}
where ${\bf y} = (y_1, \ldots, y_n)$ denotes the two-samples stacked into a single data vector of length $n = (n_1 + n_2)$. Let $\pi_0(\bm{\theta}_0)$ denote the prior distributions for the parameters under the null hypothesis. We use the usual right Haar prior for the mean and standard deviation:
\begin{equation}
	\pi_0(\bm{\theta}_0) = (\Omega \sigma)^{-1} 
\end{equation}
defined over some suitable parameter range $\Omega$ implying a location and scale invariant distribution of the mean and standard deviation, respectively. The determinant of the expected Fisher information matrix for the normal distribution is $|J(\bm{\theta}_0)| = 2n^2/\sigma^4$ leading to the MML87 codelength:
\begin{align*}
	\mathcal{I}_0({\bf y}, \bm{\theta}_0) &= -\log \ell_0(\bm{\theta}_0) + \frac{1}{2} \log |J(\bm{\theta}_0)| - \log \pi_0(\bm{\theta}_0) + \log \kappa_2 + 1 \\
%
%
&= 	\frac{1}{2 \sigma^2} \sum_{i=1}^{n} \left(y_{i} - \mu\right)^2 + (n-1) \log (\sigma ) + \frac{1}{2} \log \left( 2^{n+1} \pi^n  \left(n e \kappa _2 \Omega \right)^2\right)
\end{align*}
where $\kappa_2 = 5/36/\sqrt{3} \approx 0.0802$. MML estimates that minimize the codelength $\mathcal{I}_0({\bf y}, \bm{\theta}_0)$ correspond to the usual sample mean and the unbiased sample variance estimates
\begin{equation}
\hat{\mu}({\bf y}) = \frac{1}{n} \sum_{i=1}^n y_i = \bar{\bf y}, \quad \hat{\sigma}^2({\bf y}) = \frac{1}{n-1} \sum_{i=1}^n (y_i - \bar{\bf y})^2 .
\end{equation}
The minimum MML87 codelength for the data under the null hypothesis is then
\begin{equation}
	\mathcal{I}_0({\bf y}) = \frac{n-1}{2} \left( 1 + \log \hat{\sigma}^2({\bf y}) \right) + \frac{1}{2} \log \left( 2^{n+1} \pi^n  \left(n e \kappa _2 \Omega \right)^2\right)
\end{equation}
We observe that the MML principle allows for automatic parameter estimation and model selection, as long as we can compute codelengths of the candidate models. Next, we derive the MML87 codelength for the alternative hypothesis where the two means are different at population level ($\delta \neq 0$).

Recall that $\bm{\theta}_1 = \{\mu, \sigma, \delta\} \in (\mathbb{R}^2 \times \mathbb{R}_+) \equiv \Theta_1$ denotes the parameter set of the alternative hypothesis, which compared to the null model also includes the effect size parameter $\delta$. The negative log-likelihood of the data ${\bf y}$ under the alternative hypothesis, $-\log \ell_1({\bf y} | \bm{\theta}_1)$, is
\begin{equation}
\label{eqn:h1:nll}
	\frac{n}{2} \log(2 \pi \sigma^2) + \frac{1}{2 \sigma^2} \sum_{i=1}^{n_1} \left(y_{1i} - \mu - \frac{\sigma \delta}{2}\right)^2 + \frac{1}{2 \sigma^2} \sum_{j=1}^{n_2} \left(y_{2j} - \mu + \frac{\sigma \delta}{2}\right)^2
\end{equation}
where, as before, $n = (n_1+n_2)$ is the total sample size. Let
\begin{equation}
S_j = \sum_{i=1}^{n_j} y_{ji}, \quad S^2 = \sum_{j=1}^2 \sum_{i=1}^{n_j} y_{ji}^2 = {\bf y}^\prime {\bf y}, \quad j = 1,2 
\end{equation} 
denote the sufficient statistics. For reference, the maximum likelihood estimates of the model parameters are easily shown to be
\begin{align*}
\hat{\mu}_{\rm ML}({\bf y}) &= \frac{1}{2} \left(\frac{S_1}{n_1}+\frac{S_2}{n_2}\right) =  \frac{1}{2} \left(\bar{\bf y}_1 + \bar{\bf y}_2\right), \\
\hat{\sigma}_{\rm ML}^2({\bf y}) &= \frac{1}{n} \left( S^2 - \left(\frac{S_1^2}{n_1}+\frac{S_2^2}{n_2} \right) \right) =  \frac{1}{n} \left( S^2 - \left( n_1 \bar{\bf y}_1^2 + n_2 \bar{\bf y}_2^2\right) \right), \\
	%
\hat{\delta}_{\rm ML}({\bf y}) &= \frac{1}{\hat{\sigma}_{\rm ML}} \left(\frac{S_1}{n_1}-\frac{S_2}{n_2}\right) = \frac{1}{\hat{\sigma}_{\rm ML}} \left(\bar{\bf y}_1 - \bar{\bf y}_2\right).
\end{align*}
Next we derive the MML87 codelength for the model and the corresponding MML87 estimates. For the MML codelength of the data under the alternative hypothesis, we again require a prior distribution for all the model parameters and the determinant of the expected Fisher information matrix. Following~\cite{GronauEtAl19}, the prior distribution for the parameters $\bm{\theta}_1$ is chosen to be
\begin{equation}
\label{eqn:h1:prior}
	\pi_1(\bm{\theta}_1) = \pi_0(\bm{\theta}_0) \pi(\delta), \quad \pi(\delta) = \frac{1}{\gamma_{\delta}} T_{\kappa} \left( \frac{\delta - \mu_{\delta}}{\gamma_{\delta}}\right)
\end{equation}
where $\pi(\delta)$ is a student $t$ distribution with $\kappa > 0$ degrees of freedom, and location and scale hyperparameters $\mu_{\delta}$ and $\gamma_{\delta} > 0$, respectively. This prior density for the effect size is a symmetric, heavy-tailed flexible distribution allowing experts to incorporate prior knowledge via the hyperparameters $(\mu_{\delta}, \gamma_{\delta})$. Note that the prior distribution for the mean and standard deviation parameters is the same under the null and alternative hypotheses, which implies that the choice of the range of parameters $\Omega$ will have no effect on MML inference as it contributes equally to the codelengths of both hypotheses. The determinant of the expected Fisher information matrix under the alternative hypothesis is 
\begin{equation}
\label{eqn:h1:fisher}
	| J(\bm{\theta}_1) | = \frac{2 n_1 n_2 (n_1+n_2)}{\sigma^4} .
\end{equation}
Substituting (\ref{eqn:h1:nll}), (\ref{eqn:h1:prior}) and (\ref{eqn:h1:fisher}) into (\ref{eqn:mml87:codelength}) yields the MML87 codelength $\mathcal{I}({\bf y}, \bm{\theta}_1)$ of the data under the alternative hypothesis model. Under this choice of the prior distribution, MML87 estimates of the parameters $\bm{\theta}_1$ are unavailable in closed form and must be obtained via numerical optimisation
\begin{equation}
	\hat{ \bm{\theta} }_1({\bf y}) = \argmin_{\bm{\theta}_1 \in \Theta_1} \mathcal{I}({\bf y}, \bm{\theta}_1)	.
\end{equation}
To perform a MML $t$-test, we compute the codelength of the null model $\mathcal{I}({\bf y}, \hat{\bm{\theta}}_0)$ and the alternative model $\mathcal{I}({\bf y}, \hat{\bm{\theta}}_1)$, with the preferred model being the one with the shorter codelength. As with standard Bayesian analysis, a codelength difference of about $2.3$ nits or more indicates substantial preference of the model with the smaller codelength. Recall that we can interpret the difference in MML codelengths as the log posterior probability in favour of the model with the shorter codelength.

\begin{figure*}[t]
\begin{center}
\subfigure[$n_1=n_2=5$]{
   \includegraphics[width=6.0cm]{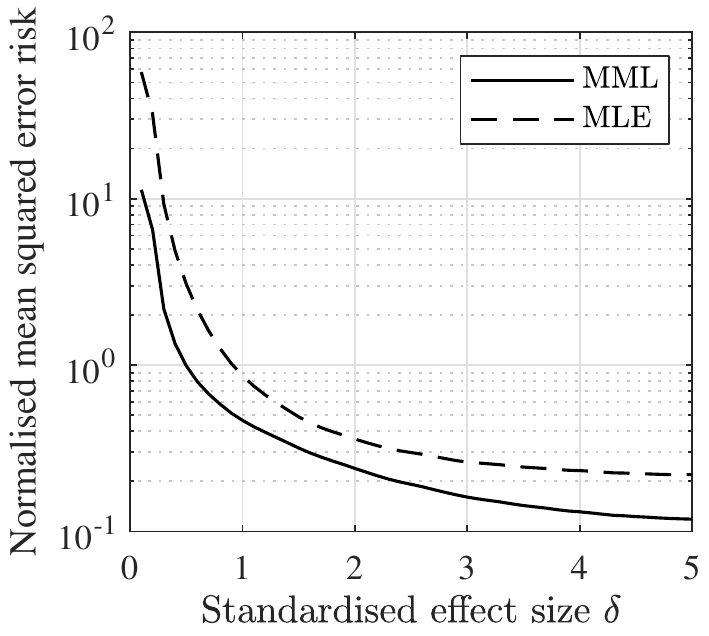}
}%
\subfigure[$n_1=n_2=10$]{
   \includegraphics[width=6.0cm]{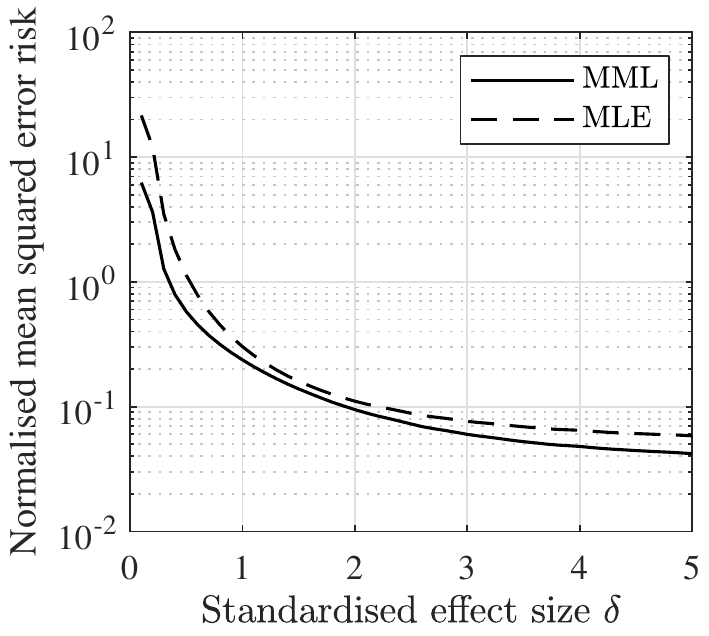}
} \\
\end{center}
\caption{Normalised mean square error for the maximum likelihood and MML estimates of $\delta$ with $n_1=n_2=5$ and $n_1=n_2=10$ data samples averaged over $10^5$ simulations. In all cases, the mean and standard deviation of the data generating model were $\mu=0, \sigma = 1$. \label{fig:delta:nmse}}
\end{figure*}

A brief simulation was performed to compare the behaviour of the MML87 estimate of the standardised effect size $\delta$ to the ML estimate. Data with sample sizes $(n_1, n_2) \in \{5, 10, 20, 50\}$ was generated from the model $(\mu = 0, \sigma = 1)$ with $0.1 \leq \delta \leq 5$ and both ML and MML87 were asked to nominate an estimate of $\delta$. The experiment was repeated for $10^5$ iterations and the estimates were compared using the normalized mean squared error metric
\begin{equation}
	\text{NMSE} = \frac{ (\delta - \hat{\delta}({\bf y}))^2 }{\delta} ,
\end{equation}
averaged over all $10^5$ iterations for each experiment. Results for sample sizes $n_1 = n_2 = 5$ and $n_1 = n_2 = 10$ are shown in Figure~\ref{fig:delta:nmse}. Behaviour of the ML and MML87 estimates for moderate to large sample sizes was indistinguishable, as predicted by asymptotic analysis. When the sample size was small, ML tends to overestimate $\delta$ and underestimate $\sigma$. This is in contrast to MML87 which tends to underestimate $\delta$ and is more conservative.

In terms of model (hypothesis) selection, unlike the frequentist methodology, neither the Bayes factor nor MML approach are designed to explicitly control type I error. Empirically, we observe that both the Bayes factor derived by Gronau et al~\cite{GronauEtAl19} and the MML87 $t$-test derived in Section~\ref{sec:mml:ttest} control type I error rate at approximately 0.10 for all $0.1 < \delta < 2$, if the alternative model is chosen whenever $\mathcal{I}_1(y) < \mathcal{I}_0(y)$, or equivalently, whenever the Bayes factor BF$_{10} > 1$. To control the type I error rate at $0.05$, we would need to select the alternative model for all BF$_{10} > 1.87$ or whenever $(\mathcal{I}_1(y) + \log(1.65)) < \mathcal{I}_0(y)$. Under this setup, all three methods were virtually indistinguishable in terms of type I error rate for all $0.1 < \delta < 2$.
\section{Minimum message length testing of the correlation coefficient}
\label{sec:mml:corr}
Suppose that a pair of random variables $(Y_1, Y_2)$ follows a bivariate normal distribution with means $(\mu_1, \mu_2)$, variances $(\sigma^2_1, \sigma^2_2)$ and correlation coefficient $\rho \in (0,1)$. 
We have $n$ observations ${\bf y} = \{(y_{1,i}, y_{2,i})\}$ ($i=1,\ldots,n$) and wish to test the hypothesis
\begin{equation*}
    H_0: \rho = \rho_0 \quad \text{versus} \quad H_1: \rho \neq \rho_0
\end{equation*}
where $\rho_0 \in (0,1)$ is a fixed, user-specified value. Let $\bm{\theta}_0 = \{\mu_1,\mu_2,\sigma_1,\sigma_2\} \in \mathbb{R}^2 \times \mathbb{R}^2_+ \equiv \Theta_0$ and $\bm{\theta}_1 = \{\mu_1,\mu_2,\sigma_1,\sigma_2,\rho\} \in \mathbb{R}^2 \times \mathbb{R}^2_+ \times (0,1) \equiv \Theta_1$ denote the parameter sets under the null and alternative hypothesis models, respectively. A large number of frequentist as well as Bayesian procedures exist for this hypothesis testing problem. A good summary of the recent literature is in \cite{PengWang21}, which also introduces a new Bayesian procedure for testing the correlation coefficient under divergence-based priors~\cite{BayarriGarciaDonato08}.

Under the null model, the negative log-likelihood of the data ${\bf y}$ is
\begin{equation}
-\log \ell_0(\bm{\theta}_0) = n \log(2\pi) + n \log(\sigma_1 \sigma_2) + \frac{n}{2} \log(1 - \rho_0^2) + \frac{ 1 }{2 \left(1-\rho_0^2\right)}  \sum_{i=1}^n Q_i({\bf y},\bm{\theta}_0)
\end{equation}
where
\begin{equation}
Q_i({\bf y},\bm{\theta}) = \frac{\left(y_{1,i}-\mu _1\right){}^2}{\sigma _1^2}-\frac{2 \rho  \left(y_{1,i}-\mu _1\right) \left(y_{2,i}-\mu _2\right)}{\sigma _1 \sigma _2}+\frac{\left(y_{2,i}-\mu _2\right){}^2}{\sigma _2^2} .
\end{equation}
Let
\begin{equation}
\bar{y}_j = \frac{1}{n} \sum_{i=1}^n y_{j,i},\quad s_j^2 = \frac{1}{n} \sum_{i=1}^n (y_{j,i} - \bar{y}_j)^2, \quad r = \frac{1}{n s_1 s_2} \sum_{i=1}^n (y_{1,i} - \bar{y}_1)(y_{2,i} - \bar{y}_2),
\end{equation}
for j = 1,2 denote the usual sufficient statistics. Maximum likelihood estimates of the model parameters are
\begin{equation}
    \hat{\mu}_j({\bf y})_{\text{ML}} = \bar{y}_j, \quad \hat{\sigma}_j^2({\bf y})_{\text{ML}} = s_j^2 \left(\frac{  1-\rho_0 \,  r}{1-\rho_0 ^2} \right), \quad (j=1,2) .
\end{equation}
For the MML codelength, we again use the right Haar prior for the means and standard deviations
\begin{equation}
    \pi_0(\bm{\theta}_0) = (\Omega_0 \sigma_1 \sigma_2)^{-1}
\end{equation}
where $\Omega_0$ is a normalisation constant. The determinant of the expected Fisher information matrix for a bivariate normal distribution with a fixed (known) correlation coefficient $\rho_0$ is 
\begin{equation}
|J(\bm{\theta}_0)| = \frac{4 n^4}{\left(1 - \rho _0^2\right){}^2 \sigma _1^4 \sigma _2^4}    
\end{equation}
and the MML87 codelength for the data ${\bf y}$ is
\begin{align*}
	\mathcal{I}_0({\bf y}, \bm{\theta}_0) &= -\log \ell_0(\bm{\theta}_0) + \frac{1}{2} \log |J(\bm{\theta}_0)| - \log \pi_0(\bm{\theta}_0) + 2 (1 + \log \kappa_4) \\
&= (n-1) \log(\sigma_1 \sigma_2) + \frac{n-2}{2} \log(1 - \rho_0^2) + \frac{ 1 }{2 \left(1-\rho_0^2\right)}  \sum_{i=1}^n Q_i({\bf y},\bm{\theta}_0) \\
& + \log \left(\Omega _0 \pi ^n 2^{n+1} (\kappa \, n)^2\right) + 2
\end{align*}
where $\kappa_4 \approx 0.076603$. MML estimates that minimise the codelength under the null hypothesis are
\begin{equation}
    \hat{\mu}_j({\bf y}) = \bar{y}_j, \quad \hat{\sigma}_j^2({\bf y}) = s_j^2 \left(\frac{ n (1-\rho_0 \,  r)}{(n-1) \left(1-\rho_0 ^2\right)} \right), \quad (j=1,2) .
\end{equation}
We observe that the MML estimates of the mean parameters are the same as the usual maximum likelihood estimates. However, MML estimates of the variances include the extra term $n/(n-1)$. If the null hypothesis corresponds to the no correlation model ($\rho_0 = 0)$, MML estimates of the variances are unbiased unlike the corresponding maximum likelihood estimates.

\begin{figure*}[tbh]
\begin{center}
\subfigure[$n=20$]{
   \includegraphics[height=6cm]{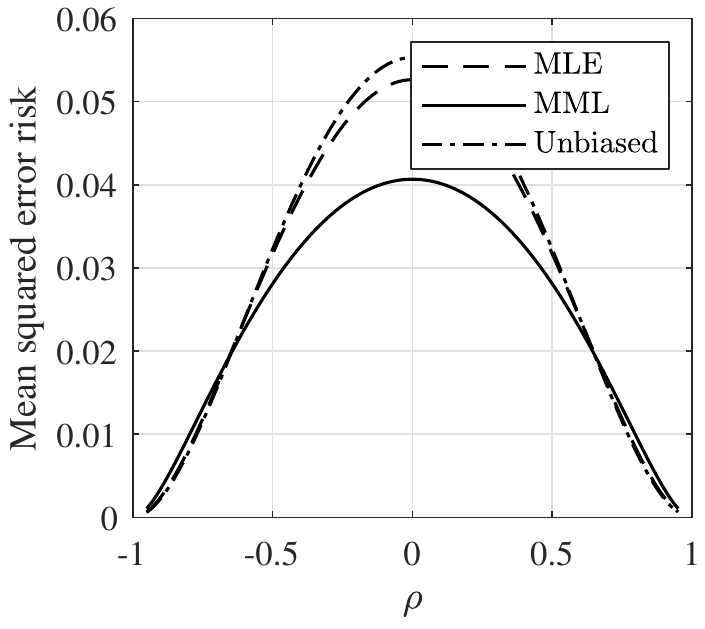}
}%
\subfigure[$n=50$]{
   \includegraphics[height=6cm]{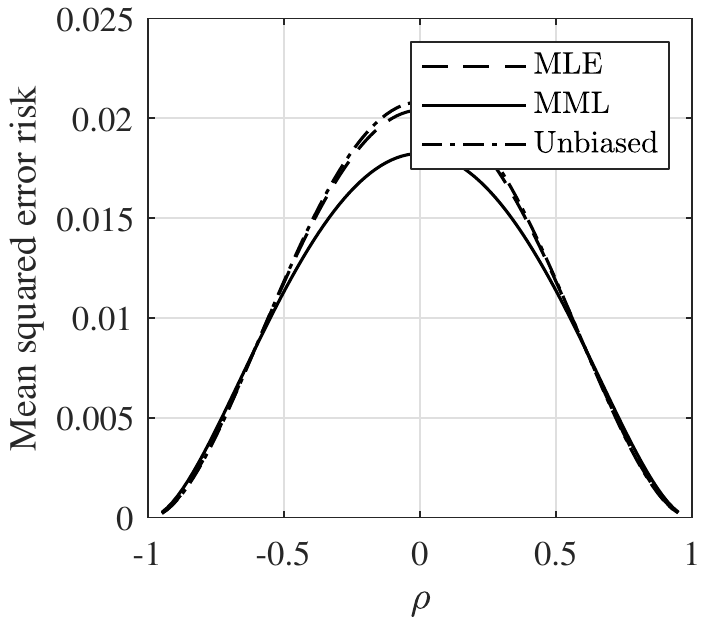}
} \\
\end{center}
\caption{Expected mean squared error for the MML, maximum likelihood and the unbiased minimum variance estimators of the correlation parameter $\rho$. \label{fig:mserisk:rho}}
\end{figure*}

Under the alternative hypothesis, the correlation coefficient $\rho$ is unknown and must be estimated from the data along with the means and variances. The negative log-likelihood of the data ${\bf y}$ is now
\begin{equation}
-\log \ell_1(\bm{\theta}_1) = n \log(2\pi) + n \log(\sigma_1 \sigma_2) + \frac{n}{2} \log(1 - \rho^2) + \frac{ 1 }{2 \left(1-\rho^2\right)}  \sum_{i=1}^n Q_i({\bf y},\bm{\theta}_1)
\end{equation}
Maximum likelihood estimates of the parameters are
\begin{equation}
        \hat{\mu}_j({\bf y})_{\text{ML}} = \bar{y}_j, \quad \hat{\sigma}_j^2({\bf y})_{\text{ML}} = s_j^2, \quad \hat{\rho}({\bf y})_{\text{ML}} = r, \quad (j=1,2) .
\end{equation}
We use the same right Haar prior for the means and standard deviations and opt for the uniform prior distribution over the correlation coefficient leading to
\begin{equation}
    \pi_1(\bm{\theta}_1) = (\Omega_1 \sigma_1 \sigma_2)^{-1}
\end{equation}
where $\Omega_1$ is a normalisation constant. The determinant of the expected Fisher information matrix for a bivariate normal distribution with an unknown correlation coefficient is 
\begin{equation}
|J(\bm{\theta}_1)| =  \frac{4 n^5}{\left(1 - \rho ^2\right)^4 \sigma _1^4 \sigma _2^4} .
\end{equation}
MML codelength of the alternative hypothesis is 
\begin{align*}
	\mathcal{I}_1({\bf y}, \bm{\theta}_1) &= -\log \ell_1(\bm{\theta}_1) + \frac{1}{2} \log |J(\bm{\theta}_1)| - \log \pi_1(\bm{\theta}_1) + \frac{5}{2} (1 + \log \kappa_5) \\
&= (n-1) \log(\sigma_1 \sigma_2) + \frac{n-4}{2} \log(1 - \rho^2) + \frac{ 1 }{2 \left(1-\rho^2\right)}  \sum_{i=1}^n Q_i({\bf y},\bm{\theta}_1) \\
& + \log \left( \Omega _1 \pi ^n 2^{n+1} (n \, \kappa _5)^{5/2} \right) + \frac{5}{2}
\end{align*}
where $\kappa_5 \approx 0.075625$. MML estimates that minimise the codelength are
\begin{align*}
\hat{\mu}_j({\bf y}) &= \bar{y}_j, \\
\hat{\sigma}^2_j ({\bf y}) &= s_j^2 \left(\frac{n  (n + 2 + \sqrt{(n+2)^2 - 12 r^2 (n-1)} )} {2(n+2)(n-1)} \right) = s_j^2 \left( \frac{n (n-3 \hat{\rho}  r+2)}{(n-1) (n+2)} \right) \\
\hat{\rho}({\bf y}) &= \frac{n + 2 - \sqrt{(n+2)^2 - 12 r^2 (n-1)} }{6 r} = \frac{(n+2)}{3 r} \left(1-\left(1-\frac{1}{n}\right) \left(\frac{\hat{\sigma}_j}{s_j}\right)^2 \right) .
\\
&= r \left( \frac{ n-1 } {n+2} \right) + O\left(n^{-1}\right)
\end{align*}
for $j= 1,2$. It is easy to show that the MML estimate of the correlation parameter is always smaller in magnitude compared to the sample correlation coefficient; ie, $|\hat{\rho}({\bf y})| < |r|$. Figure~\ref{fig:mserisk:rho} shows the mean squared error (MSE) risk for $n \in \{20, 50\}$ for the MML, maximum likelihood and the unique minimum variance unbiased estimator~\cite{OlkinPratt58}
\begin{equation*}
    \hat{\rho}({\bf y})_{\text{Unbiased}} = r \, _2F_1\left(\frac{1}{2},\frac{1}{2};\frac{n-1}{2};1-r^2\right)
\end{equation*}
of the correlation parameter $\rho$; here, $_2 F_1(\cdot)$ is Gaussian hypergeometric function. When $n=20$, the MML estimate dominates the maximum likelihood estimate in terms of MSE risk for all $|\rho| < 0.64$. As $|\rho| \to 1$, all three estimates are virtually indistinguishable. When $n=50$, the MML estimate again dominates the maximum likelihood estimate for all $|\rho| < 0.6$. 

\begin{table*}[tbph]
\scriptsize
\begin{center}
\begin{tabular}{ccccccccccccc} 
\toprule
$\rho_0$ & $\rho$ & $n$ & \multicolumn{7}{c}{Relative frequency of rejecting $H_0: \rho = \rho_0$} & & \multicolumn{2}{c}{KL Divergence} \\
    &     &     & $\text{DB}^S$ & $\text{DB}^M$ & Fisher Z & HP & Mud & Rud & MML & ~ & MLE & MML \\
\cmidrule{1-13}
\multirow{8}{*}{-0.30} & \multirow{2}{*}{-0.75} &  15 & 0.730  & {\bf 0.812}  & 0.704  & 0.730  & 0.702  & 0.679  & 0.570  &  & {\bf 0.408} & 0.414 \\ 
 & & 30 & 0.921  & {\bf 0.950}  & 0.936  & 0.938  & 0.936  & 0.932  & 0.890  &  & {\bf 0.129} & 0.145 \\ 
 & \multirow{2}{*}{-0.30} &  15 & 0.073  & 0.124  & 0.060  & 0.068  & 0.059  & 0.058  & {\bf 0.050}  &  & 0.262 & {\bf 0.230} \\ 
 & & 30 & 0.043  & 0.081  & 0.046  & 0.050  & 0.046  & 0.046  & {\bf 0.035}  &  & 0.132 & {\bf 0.124} \\ 
 & \multirow{2}{*}{0.00} &  15 & 0.236  & {\bf 0.329}  & 0.184  & 0.206  & 0.178  & 0.193  & 0.245  &  & 0.229 & {\bf 0.190} \\ 
 & & 30 & 0.354  & {\bf 0.465}  & 0.371  & 0.382  & 0.371  & 0.382  & 0.358  &  & 0.092 & {\bf 0.083} \\ 
 & \multirow{2}{*}{0.50} &  15 & 0.908  & {\bf 0.934}  & 0.883  & 0.893  & 0.881  & 0.887  & 0.904  &  & 0.278 & {\bf 0.223} \\ 
 & & 30 & 0.995  & {\bf 0.998}  & 0.995  & 0.996  & 0.995  & 0.995  & 0.994  &  & 0.105 & {\bf 0.097} \\ 
\vspace{-2mm} \\ 
\cmidrule{2-13}
\vspace{-2mm} \\ 
\multirow{8}{*}{0.00} & \multirow{2}{*}{-0.75} &  15 & 0.969  & {\bf 0.988}  & 0.953  & 0.960  & 0.950  & 0.950  & 0.938  &  & 0.298 & {\bf 0.252} \\ 
 & & 30 & 0.998  & 0.998  & 0.998  & 0.998  & 0.998  & 0.998  & {\bf 0.999}  &  & 0.106 & {\bf 0.099} \\ 
 & \multirow{2}{*}{-0.30} &  15 & 0.262  & {\bf 0.362}  & 0.207  & 0.224  & 0.204  & 0.204  & 0.198  &  & 0.277 & {\bf 0.237} \\ 
 & & 30 & 0.352  & {\bf 0.476}  & 0.356  & 0.368  & 0.356  & 0.356  & 0.322  &  & 0.126 & {\bf 0.118} \\ 
 & \multirow{2}{*}{0.00} &  15 & 0.074  & 0.130  & 0.055  & 0.060  & 0.055  & 0.055  & {\bf 0.051}  &  & 0.217 & {\bf 0.189} \\ 
 & & 30 & 0.050  & 0.106  & 0.053  & 0.057  & 0.053  & 0.053  & {\bf 0.037}  &  & 0.087 & {\bf 0.080} \\ 
 & \multirow{2}{*}{0.50} &  15 & 0.613  & {\bf 0.700}  & 0.559  & 0.580  & 0.554  & 0.548  & 0.508  &  & 0.330 & {\bf 0.277} \\ 
 & & 30 & 0.809  & {\bf 0.893}  & 0.820  & 0.057  & 0.053  & 0.053  & 0.791  &  & 0.126 & {\bf 0.120} \\ 
\vspace{-2mm} \\ 
\cmidrule{2-13}
\vspace{-2mm} \\ 
\multirow{8}{*}{0.70} & \multirow{2}{*}{-0.75} &  15 & {\bf 1.000}  & 1.000  & 1.000  & 1.000  & 1.000  & 1.000  & 1.000  &  & 0.273 & {\bf 0.232} \\ 
 & & 30 & {\bf 1.000}  & 1.000  & 1.000  & 1.000  & 1.000  & 1.000  & 1.000  &  & 0.106 & {\bf 0.098} \\ 
 & \multirow{2}{*}{-0.30} &  15 & 0.984  & 0.987  & 0.985  & 0.986  & 0.985  & 0.986  & {\bf 0.992}  &  & 0.264 & {\bf 0.221} \\ 
 & & 30 & {\bf 1.000}  & 1.000  & 1.000  & 1.000  & 1.000  & 1.000  & 1.000  &  & 0.105 & {\bf 0.096} \\ 
 & \multirow{2}{*}{0.00} &  15 & 0.834  & 0.875  & 0.850  & 0.865  & 0.849  & 0.867  & {\bf 0.910}  &  & 0.242 & {\bf 0.207} \\ 
 & & 30 & 0.992  & 0.998  & 0.998  & 0.998  & 0.998  & {\bf 0.999}  & 0.993  &  & 0.104 & {\bf 0.095} \\ 
 & \multirow{2}{*}{0.50} &  15 & 0.131  & 0.182  & 0.150  & 0.167  & 0.149  & 0.173  & {\bf 0.258}  &  & 0.328 & {\bf 0.287} \\ 
 & & 30 & 0.275  & 0.397  & 0.403  & 0.412  & 0.401  & {\bf 0.417}  & 0.354  &  & {\bf 0.176} & 0.178 \\ 
\vspace{-3mm} \\ 
\bottomrule
\vspace{+1mm}
\end{tabular}
\caption{Relative frequency of rejecting the null hypothesis and Kullback--Leibler (KL) divergence from the data generating model for maximum likelihood (MLE) and minimum message length (MML) estimates. \label{tab:results:corr}}
\end{center}
\end{table*}

\section{Discussion}
\label{sec:sim}
An advantage of MML over alternative approaches is that it allows conceptually straightforward comparison of models with different structures and parameters as long as we can compute their codelengths. The codelength is measured in the unit of information (e.g., bits) and is therefore a universal yardstick for model comparison. It is then easy to extend the MML $t$-test to, for example, handle the case of unequal variances in the two groups, a version of the so-called Behrens--Fisher problem. The setup is now 
\begin{equation}
	Y_{1i} \sim \mathcal{N}\left(\mu_1, \sigma_1^2\right), \quad Y_{2j} \sim \mathcal{N}\left(\mu_2, \sigma_2^2\right) ,
\end{equation}
for $i=1,\ldots,n_1$ and $j = 1, \ldots, n_2$ and the task is, given data ${\bf y}$, to determine whether the two means are equal $\mu_1 = \mu_2$. MML87 codelengths for this setup were derived in \cite{MakalicSchmidt13} and can be compared to the codelengths of the null and alternative hypotheses under the equal variance assumption derived in Section~\ref{sec:mml:ttest} to discriminate between the four candidate models. Of course, we can also include other models in the comparison, as long as we can compute the corresponding codelengths.

\subsection{Minimum description length}
Minimum description length (MDL)~\cite{Rissanen78,Rissanen96,RissanenRoos07a,Rissanen09b,GrunwaldRoos19,BruniEtAl22} is a closely related approach to inductive inference that was developed around the same time as MML by Rissanen and colleagues and has gone through various refinements. MDL seeks a model class $\mathcal{M}$ that results in the shortest codelength of the data (ie, a model that best compresses the data), where a model class is defined as a set of parametric distributions indexed by parameter $\bm{\theta}$:
\begin{equation*}
    \mathcal{M} = \{ p(\cdot | \bm{\theta}): \bm{\theta} \in \Theta\} .
\end{equation*}
A popular version of MDL is the general normalized maximum likelihood (NML) code whose codelength is given by
\begin{equation}
    -\log p_{\text{NML}}({\bf y} | v, \mathcal{M}) = \underbrace{-\log \left[\max_{\bm{\theta} \in \Theta}  p({\bf y} | \bm{\theta}({\bf y}), \mathcal{M}) \,  v(\bm{\theta}) \right] }_{\rm model\; fit} + \underbrace{\log \left[\sum_{{\bf x}} \max_{\bm{\theta} \in \Theta} p({\bf x} | \bm{\theta}({\rm x}), \mathcal{M}) \, v(\bm{\theta}) \right]}_{\rm parametric\; complexity}
\end{equation}
where $v: \Theta \to [0, \infty)$ is a user-specified function, referred to as the \emph{luckiness} function in MDL terminology, and the sum (integral) is over the entire data space in the case of discrete (continuous) data. This is the general NML codelength for data ${\bf y}$ with respect to some model class $\mathcal{M}$ parameterised by models $\bm{\theta} \in \Theta \in \mathcal{M}$. We observe that the NML codelength is a sum of two terms:
\begin{enumerate}
    \item the (penalized) negative log-likelihood of the data evaluated at the maximum, and
    \item the so-called parametric complexity of the model class $\mathcal{M}$ which measures how well models $\bm{\theta} \in \mathcal{M}$ approximate random data sequences. 
\end{enumerate}
The first term in the NML codelength represents how well the model $\hat{\theta} \in \mathcal{M}$ fits the observed data with smaller codelengths indicating better data fit. In contrast, a low (high) parametric complexity implies that a few (many) data sequences can be modelled sufficiently well by models in class $\mathcal{M}$. The parametric complexity denotes the logarithm of the number of distinguishable distributions in the model class~\cite{Balasubramanian05} and is similar in spirit to the assertion in the MML code, accounting for the complexity of the model class from which the maximum likelihood estimate is derived. A detailed discussion of the NML distribution and the corresponding optimality properties is available in~\cite{Rissanen01,Rissanen07,GrunwaldRoos19}.


Originally, Risannen advocated the use of the NML distribution for $v(\theta) = 1$ only. Unfortunately, the parametric complexity in this case is easily shown to be infinite for many model classes rendering a straightforward application of the NML codelength impossible. Subsequently, Rissanen and colleagues  suggested other forms of the luckiness function as well as variations of the NML code such as the restricted approximate normalized maximum likelihood (ANML), the two-part ANML or the objective Bayesian code, among others~\cite{RooijGrunwald06,GrunwaldRoos19}. \\

\noindent {\bf Example:} Assuming $v(\theta) = 1$, the NML codelength for the binomial distribution is
\begin{equation}
    -\log p_{\text{NML}}(y | \mathcal{M}) = -\log \binom{n}{y} \hat{\theta}^y (1-\hat{\theta})^{n-y}+ \log \mathcal{C}(n)
\end{equation}
where
\begin{equation}
    \mathcal{C}(n) = \sum_{x = 0}^n \binom{n}{x}  \left( \frac{x}{n}\right)^{x} \left(\frac{n - x}{n}\right)^{n - x} 
\end{equation}
and $\hat{\theta}(y) = (y/n)$ is the maximum likelihood estimate of the success probability. An efficient linear-time algorithm to compute the parametric complexity of a general multinomial distribution is available in~\cite{KontkanenMyllymaki07b}. As an example, if $n = 10$ and $y = 3$, the maximum likelihood estimate is $\hat{\theta}(y) = 0.3$, the parametric complexity is $\log \mathcal{C}(n) \approx 2.22$ bits and the total NML codelength is approximately 4.13 bits. 

For the original NML distribution with $v(\theta) = 1$, Rissanen~\cite{Rissanen96} showed that 
\begin{equation}
    -\log p_{\text{NML}}({\bf y} | \mathcal{M}) = -\log p({\bf y} | \hat{\bm{\theta}}({\bf y}), \mathcal{M}) + \log \int_\Theta \sqrt{|J_1(\bm{\theta}) | } + \frac{p}{2} \log \left( \frac{n}{2\pi} \right) + o(1)
\end{equation}
where $J_1(\cdot)$ is the per-sample Fisher information matrix. Additionally, Rissanen proved that the NML codelength reduces to the well-known Bayesian information criterion (BIC) in the limit as the sample size $n \to \infty$ and demonstrated that the asymptotic approximation is accurate for large sample sizes. Other, somewhat shaper, approximations to this NML codelength exist, see for example, Mera et al.~\cite{MeraEtAl2020}. Rissanen's asymptotic expansion was recently generalized to include arbitrary luckiness functions and a novel methodology for evaluation of the general, non-asymptotic NML codelength for exponential family models~\cite{SuzukiYamanishi18}.


Although there are many similarities between MML and MDL, there exist some important differences in their inherent inference philosophies. Unlike MML, MDL is non-Bayesian and does not allow for the use of prior information in inference. While MML provides new means of parameter estimation, MDL was originally based on the maximum likelihood estimator and aimed to discover the best model class, rather than the best single model, for a given data set. However, the recent inclusion of luckiness functions in the NML code allows for new MDL-based penalized maximum likelihood estimation procedures. Lastly, the NML codelength is not a two-part code as Strict MML and is instead derived to minimize the worst-case excess codelength relative to an ideal code. For a more detailed discussion of MML and MDL, we recommend~\cite{BaxterOliver94} and \cite{Wallace05} (pp. 413--415).

\bibliographystyle{amsplain} 
\bibliography{Bibliography/bibliography} 
\end{document}